\documentclass[11pt]{article}
\usepackage{latexsym}
\usepackage{amssymb}
\usepackage{amsmath,amsthm}
\usepackage{graphicx}
\usepackage{fullpage}
\usepackage{authblk}
\usepackage[hidelinks]{hyperref}
\usepackage{enumitem}
\usepackage[longend,noresetcount,boxed]{algorithm2e}
\usepackage{tikz}
\usetikzlibrary{positioning}
\usepackage{xcolor}
\usepackage{colortbl}
\usepackage{ifthen}
\usepackage{soul}

\usepackage{anysize}
\usepackage{microtype}
\marginsize{1in}{1in}{1in}{1in}
\begin{document}
\DontPrintSemicolon
\LinesNotNumbered
\SetAlCapSkip{8pt}
\def\qed{\hfill$\Box$}
\def\Nset{\mathbb{N}}
\def\Ascr{\mathcal{A}}
\def\Bscr{\mathcal{B}}
\def\Cscr{\mathcal{C}}
\def\Dscr{\mathcal{D}}
\def\Escr{\mathcal{E}}
\def\Fscr{\mathcal{F}}
\def\Hscr{\mathcal{H}}
\def\Iscr{\mathcal{I}}
\def\Jscr{\mathcal{J}}
\def\Mscr{\mathcal{M}}
\def\Nscr{\mathcal{N}}
\def\Pscr{\mathcal{P}}
\def\Qscr{\mathcal{Q}}
\def\Rscr{\mathcal{R}}
\def\Sscr{\mathcal{S}}
\def\Wscr{\mathcal{W}}
\def\Xscr{\mathcal{X}}
\def\cupp{\stackrel{.}{\cup}}
\def\bold{\bf\boldmath}

\newcommand{\NP}{\mbox{\slshape NP}}
\newcommand{\opt}{\mbox{\scriptsize\rm OPT}}
\newcommand{\lin}{\mbox{\scriptsize\rm $OPT_{LP}$}}
\newtheorem{theorem}{Theorem}
\newtheorem{lemma}[theorem]{Lemma}
\newtheorem{corollary}[theorem]{Corollary}
\newtheorem{proposition}[theorem]{Proposition}
\newtheorem{definition}[theorem]{Definition}
\newtheorem{observation}[theorem]{Observation}
\newtheorem{claim}{Claim}
\newenvironment{proof_claim}{\noindent {\it Proof of the Claim: }}{\hspace*{\fill} $\diamond$\smallskip}

\newcommand{\E}[1]{\mathbb{E}\left[#1\right]}

\title {The Salesman's Improved Paths:  \\$3/2+1/34$ Integrality Gap and Approximation Ratio \thanks{Research initiated during the  Hausdorff Trimester Program in Combinatorial Optimization, Hausdorff Institute of Mathematics, Bonn; Partially supported by LabEx PERSYVAL-Lab (ANR 11-LABX-0025) and by a grant from the Simons Foundation (\#359525, Anke Van Zuylen). }}
\author[1]{Andr\'as Seb\H{o}}
\author[2]{Anke van Zuylen}
\affil[1]{\small CNRS, Univ. Grenoble Alpes,  Optimisation Combinatoire (G-SCOP) 
	}
\affil[2]{\small Department of Mathematics\\ College of William \& Mary\\ Williamsburg, VA }

\maketitle

	\begin{abstract}
		We give a new, strongly polynomial-time
		algorithm and improved analysis for the metric $s-t$ path TSP. 
		It finds a tour of cost  less than 1.53 times the optimum of the subtour elimination LP,  while known examples show that 1.5 is a lower bound for the integrality gap.
		
		A key new idea is the deletion of some edges of Christofides' trees, which is then accompanied by novel arguments of the analysis:   edge-deletion disconnects the trees, which are then partly reconnected by ``parity correction''. We show that the arising ``connectivity correction'' can be achieved for a minor extra cost.
		
		On the one hand this algorithm and analysis  extend previous tools such as the best-of-many Christofides algorithm.  On the other hand, powerful new tools are solicited, such as a  flow problem for analyzing the reconnection cost, and the  construction of a set of more and more restrictive spanning trees, each of which can still be found by the greedy algorithm. We show that these trees can replace the convex combination of spanning trees in the best-of-may Christofides algorithm. 
		
		These new methods lead to improving the integrality ratio and approximation guarantee below 1.53, as it is already sketched in the preliminary shortened version of this article that appeared in FOCS 2016. The algorithm and analysis have been significantly simplified in the current article, and details of proofs and explanations have been added.  
	\end{abstract}

	\newcommand{\tr}{\triangle\{s,t\}}

	\def\qed{\hfill$\Box$}
	\def\Nset{\mathbb{N}}
	\def\Ascr{\mathcal{A}}
	\def\Bscr{\mathcal{B}}
	\def\Cscr{\mathcal{C}}
	\def\Dscr{\mathcal{D}}
	\def\Escr{\mathcal{E}}
	\def\Fscr{\mathcal{F}}
	\def\Hscr{\mathcal{H}}
	\def\Iscr{\mathcal{I}}
	\def\Jscr{\mathcal{J}}
	\def\Mscr{\mathcal{M}}
	\def\Nscr{\mathcal{N}}
	\def\Pscr{\mathcal{P}}
	\def\Qscr{\mathcal{Q}}
	\def\Rscr{\mathcal{R}}
	\def\Sscr{\mathcal{S}}
	\def\Wscr{\mathcal{W}}
	\def\Xscr{\mathcal{X}}
	\def\cupp{\stackrel{.}{\cup}}
	\def\bold{\bf\boldmath}

	\maketitle

	\section{Introduction}\label{sec:i}
	In the Traveling Salesman Problem (TSP), we are given a set $V$ of $n$ ``cities'', a cost function $c: {V\choose 2}\to \mathbb{Q}_{>0}$, and the goal is to find a circuit of minimum cost that starts and ends in the same city and visits each city exactly once. This ``minimum-length Hamiltonian circuit'' problem is one of the most well-known problems of combinatorial optimization. It is not only \NP-hard to solve but also to approximate with any constant approximation ratio, and even for quite particular cost functions, since  the Hamiltonian circuit problem in $3$-regular, planar, 3-connected graphs is \NP-hard \cite{GarJT76}. For a thoughtful and entertaining account of the difficulties and successes of the TSP, see Bill Cook's book \cite{Cook11}. 
	
	A condition on the cost function that helps in theory and is often satisfied in practice is known as the triangle inequality in complete graphs. A nonnegative function satisfying this inequality is called a {\em metric}. If the cost function is a metric, we may relax the problem and allow a walk that starts and ends in the same city, and visits each city {\em at least} once; such a walk can be shortcut to a circuit which visits every city exactly once without increasing the cost.
	
	\medskip 
	Christofides~\cite{Christofides76} gave a very simple $\frac32$-approximation algorithm for the metric TSP:
	he separated the problem into finding a minimum-cost solution that is connected (a minimum-cost spanning tree) and then completing this solution to make every degree even with minimum cost.
	More precisely, 	Christofides' algorithm first finds a minimum-cost spanning tree $S$ in the complete graph with node set $V$, and edge costs given by $c$, and then adds  a {\em parity correction} to $S$ by adding a minimum-cost $T_S$-join, where
	 a $T$-join $(T\subseteq V,\,\hbox{$|T|$ is even})$ is a set of edges $J$ such that $|J\cap \delta(\{v\})|$ is odd for $v\in T$ and  even for $v\in V\setminus T$, and 
  $T_S$ is the set of nodes that have an odd number of  incident edges in $S$.

	Wolsey~\cite{Wolsey80}  (see also Cunningham \cite{Cunningham86} and Shmoys and Williamson~\cite{ShmoysW90}) observed that Christofides' algorithm actually finds a solution that can be bounded with the optimum of the well-known {\em subtour elimination} linear program (LP) that was introduced by Dantzig, Fulkerson and Johnson \cite{DantzigFJ54}:
	\begin{eqnarray}\nonumber
	& \mbox{Min} & c(x):= \sum_{e \in {V\choose 2}} c(e) x(e) \nonumber\\
	& \mbox{subject to:}  &  \sum_{e \in \delta(\{v\})} x(e) = f(\{v\}), \ \  \hbox{ for all } v \in V, \label{degreecons} \nonumber\\
	& &  \sum_{e \in \delta(U)}x(e) \geq f(U), \ \  \hbox{ for all } \emptyset \subsetneq U\subsetneq V,\, \label{subtourcons}\nonumber\\
	& & x(e) \geq 0, \ \ \ \hbox{ for all } e \in {V\choose 2}, \label{boundscons}\nonumber
	\end{eqnarray}
	where $f(U)\equiv 2$ for all sets $U$, and $\delta(U)=\{\{u,v\}\in {V\choose 2}:u\in U, v\not\in U\}$. The first set of constraints impose the condition that every city is ``both entered and left'' exactly once, and the second set of constraints, known as the {\em subtour elimination constraints}, ensure that every subset of cities is ``both entered and left'' at least once. In the rest of this article, let $x^*$ denote an optimal solution to the subtour elimination LP, and $\lin:=\lin(c):=c(x^*)$ the optimal objective value.
	
	The {\em integrality gap} is the maximum of $OPT/\lin$ over all metric cost functions $c$, where $OPT$ is the minimum of the cost of a  Hamiltonian circuit for the metric $c$. Wolsey~\cite{Wolsey80} and independently Cunningham \cite{Cunningham86} observed that $\frac{n-1}{n}x^*$ is in the spanning tree polytope, and that $x^*/2$ is in the $T$-join polyhedron for any set $T$ of even size, which implies that the integrality gap of the subtour elimination LP is at most $\frac32$.
	Despite significant effort, no improvement on the bound of 3/2 is known for either the approximation ratio or the integrality gap. 
	\medskip
	
	A relevant generalization of the (metric) TSP is the (metric) $s-t$ path TSP, where $s,t\in V$ are part of the input. The salesman starts in $s$,  ends in $t$, and needs to visit every  city once. In other words, the goal is to find a Hamiltonian path from $s$ to $t$ of minimum cost. The special case $s=t$  is the regular traveling salesman problem.  When $s\neq t$, the subtour elimination LP for the $s-t$ path TSP is as above, but now defining $f(U)=1$ if $|U\cap \{s,t\}|=1$ and $f(U)=2$ otherwise.

	To the best of our knowledge, the first relevant occurrence of $s\ne t$  is in an exercise in~\cite{JohnsonP85} concerning the case when the endpoints are not fixed, which is easily reducible to the $s=t$ case.
	Hoogeveen~\cite{Hoogeveen91} provides a Christofides-type approximation algorithm for the metric case, with an approximation ratio of $5/3$ rather than $3/2$. 
	As in Christofides' algorithm, the algorithm takes a minimum-cost spanning tree $S$, and corrects the parity of $S$. Letting $\triangle$ denote the symmetric difference operation, the node set  with the ``wrong'' degree parity is $T_S\triangle \{s,t\}$, and thus a $T_S\triangle\{s,t\}$-join is added to correct the parity of the nodes. 
	
	There had been no improvement until An, Kleinberg and Shmoys \cite{AnKS}  improved this ratio to $\frac{1+\sqrt{5}}{2}<1.618034$ with a simple algorithm and an ingenious new framework for the analysis.
	The algorithm in \cite{AnKS} is the {\em best-of-many}  version of Christofides' algorithm. It first determines a minimum-cost solution $x^*$ of the subtour elimination LP.  Then  writing $x^*$ as a convex combination of spanning trees and adding  Christofides's {\em parity correction} for each, the algorithm outputs the best of the arising solutions. The best-of-many algorithm was used by subsequent publications~~\cite{Sebo13},  \cite{Vygen16}, \cite{GottschalkV16} and is also used in the present work with some modifications.
	\medskip
	
	For the $s-t$ path TSP on ``graph metrics'', that is, cost functions that are defined as the shortest path distances on a given unweighted graph, Gao~\cite{Gao13}  proves that, for an LP solution $x^*$, there exists a tree in $E=\{i: x^*_i>0 \}$ that has exactly one edge in the so-called narrow cuts for $x^*$,   (see Section~\ref{sec:prelim} below). For such a tree, it is possible to bound the cost of the parity correction by $c(x^*)/2$, as in Wolsey's \cite{Wolsey80} or Cunningham's \cite{Cunningham86}  version of Christofides' analysis. This allows Gao to give  a very elegant proof of the approximation ratio $3/2$ for graph metrics, a result that was first shown by Seb\H{o} and Vygen~\cite{SeboV14} with a combinatorial, but more difficult proof.

	For arbitrary metrics, the best-of-many Christofides-algorithm was improved using the fruitful idea of {\em choosing the convex combination of spanning trees in a particular way} by Vygen~\cite{Vygen16}. The claimed progress in the ratio was only $0.001$, but the reassambling of trees  which participate in the convex combination by local changes is further developed by Gottschalk and Vygen~\cite{GottschalkV16}. 
	They generalize the concept of Gao-trees, and the reassambling leads to a powerful result: a convex combination using  generalized Gao-trees. 
	Although the use of local changes  in Gottschalk and Vygen's proof of the existence of the convex combination is constructive, the algorithm it implies is not a polynomial-time algorithm.  Kanstantsin Pashkovich pointed out that this convex combination can be found (in weakly polynomial time) using the ellipsoid method \cite{GottschalkV16}. Gottschalk and Vygen show that the best-of-many algorithm applied to this convex combination gives a $1.566$-approximation and integrality ratio.

	Very recently, a completely different algorithmic approach has been initiated by Traub and Vygen~\cite{TraubV18}: They propose a dynamic programming algorithm that computes a solution of cost argued to be at most $(3/2 + \varepsilon)OPT$ in time exponential as a function of the  size of the input including $\varepsilon$, but in polynomial time for fixed $\varepsilon$. Continuing this line of work with elegant new ideas, Zenklusen~\cite{Zenklusen18} finds a solution of cost at most $3/2\, OPT$ with a nice polynomial-time algorithm. However,  the conjecture that $OPT$ can be replaced here by $\lin$, remains open.

	\medskip
	
	In this paper we describe a new strongly polynomial-time algorithm for the metric $s-t$ path traveling salesman problem, and an analysis introducing novel elements. This leads to a bound on the cost of the resulting Hamiltonian path of $(3/2+1/34)\lin$, and a proof of this has been sketched in a preliminary version of this paper \cite{AA}. There is actually no example showing that the found Hamiltonian path may be larger than $3/2 \lin$,  leaving the possibility for a better bound.

	For the improved algorithm, we alter the best-of-many Christofides' algorithm by {\em deleting} certain edges from the spanning tree, in the hope that the resulting forest will be automatically reconnected during parity correction; however, this hope is not always satisfied, and whenever it is not, we also need to invest separately in {\em reconnection}. We deal with this in the algorithm by anticipating reconnection costs in the parity correcting  $T$-joins; for the analysis we use an LP to define the distribution of a random choice for the reconnecting edges in a balanced way. This LP can be solved combinatorially (unlike the subtour elimination LP) in polynomial time, using bipartite matchings or network flows. Random sampling is just a tool for an intuitive formulation of the analysis, and does not affect the deterministic nature of our algorithms.
	
	The convex combination of Gottschalk-Vygen will also play a role in our analysis; we need certain properties of the trees in order to bound the reconnection cost. We provide a new interpretation of this convex combination in terms of matroid partition, which immediately implies a polynomial-time algorithm for finding such a convex combination. This connection to matroid partition has also lead to a simpler proof of the existence of this convex combination~\cite{SSTvZ}. However, it turns out that for the algorithm presented here we only need the {\em existence} of this convex combination, with the  matroid-basis property of generalized Gao-trees: using this property, our algorithm can work with a set of spanning trees found by a greedy algorithm rather than a convex combination.

	Although the essence of the algorithm and analysis we describe in this paper is the same as those described in the preliminary version that appeared in FOCS 2016 \cite{AA}, both the algorithm and the analysis have been significantly simplified, leading to the current presentation.

	\section{Preliminaries}\label{sec:prelim}
	In this section, we introduce our notation, and give a more detailed description of the best-of-many Christofides' algorithm and results from the literature that we need for our algorithm and analysis.
	\medskip
	
	Given a finite set $V$ and a metric $c$ on $V$, 
	a  minimum-cost solution $x^*$ to the subtour elimination LP for the $s-t$ path TSP can be determined in polynomial time \cite{GLS};  we fix $E:=\{e: x^*(e)>0\}$ and set $G=(V,E)$. With a slight abuse of notation, we use the same notation for a (multi) subset of $E$ and its own incidence vector in $\mathbb{Z}_{\ge 0}^E$.   For a vector $c\in \mathbb{R}^E$ and any (multi) subset $H$ of $E$, $c(H)$ or $c(z_H)$, --- where $z_H$ is the multiplicity vector of $H$ ---  denotes the scalar product of $c$ and $z_H$, that is, $\sum_{e\in E}c(e)z_H(e)$.  When all  multiplicities are $1$, this is becoming the usual notation $c(H):=\sum_{e\in H} c(e)$.  
	For two sets $A$, $B$, let $A\triangle B:=(A\setminus B) \cup (B\setminus A)$ be the symmetric difference operation, which corresponds to the mod~$2$ sum of the incidence vectors. The operation ``$+$'' between sets means the disjoint union (sum of the multiplicity vectors).

	A multigraph is a graph where each edge has a multiplicity; the degree of a node of the multigraph is the sum of multiplicities of edges incident to a node.  A multisubgraph of a graph $G$ is a multigraph whose edges of positive multiplicity form a subgraph of $G$.
	An {\em $\{s,t\}$-tour} -- if $s=t$ a {\em tour} of $G$ -- is a multisubgraph of $G$, connected on the node-set $V$ (equivalently, a spanning connected multisubgraph) in which $s$ and $t$ have odd degree if $s\ne t$, and every other node has even degree. 
	We recall from the introduction that a $T$-join $(T\subseteq V,\,\hbox{$|T|$ is even})$ is a set of edges $J$ such that $|J\cap \delta(\{v\})|$ is odd for $v\in T$ and $|J\cap \delta(\{v\})|$ is even for $v\in V\setminus T$.
	
	\medskip
	We can write $x^*$ as a convex combination of a polynomial number, $O(|V|^2)$, of spanning trees in polynomial time (Caratheodory's theorem) ; that is,  a collection of  spanning trees $\Sscr$ and coefficients $\lambda_S>0$ for each $S\in \Sscr$ such that $\sum_{S\in \Sscr} \lambda_S=1$ and $x^*=\sum_{S\in \Sscr} \lambda_S S$ can be determined in polynomial time, see for instance~\cite{AnKS}.   
	
	The best-of-many Christofides' (BOMC) algorithm of \cite{AnKS} expresses a minimum-cost solution $x^*$ to the subtour elimination LP as a convex combination of trees, $x^*=\sum_{S\in \Sscr} \lambda_S S$, then
	computes a minimum-cost $T_S\tr$-join $J_S$ for every tree $S$ with $\lambda_S>0$, and finally outputs the minimum cost  $\{s,t\}$-tour among the tours $S+J_S$ $(S\in\Sscr$).

	The cost of the BOMC solution is at most $\sum_{S\in \Sscr} \lambda_S c(S+J_S)$, a convex combination of the costs of the constructed $\{s,t\}$-tours.  Clearly, $\sum_{S\in \Sscr}\lambda_Sc(S)=c(x^*)$; the main difficulty for the analysis of the $s-t$ path TSP version of Christofides' algorithm is the ``parity correction'' part.  The $T_S\tr$-join polyhedron is (Edmonds and Johnson~\cite{EdmondsJ73}):
	\begin{eqnarray*}
		\sum_{e\in \delta(U)}y(e)\ge 1&&\mbox{for all $U$ such that $|U\cap T_S\tr|$ is odd},\label{oddcuts}\\
		y(e) \ge 0 &&\mbox{for all }e\in E.
	\end{eqnarray*}
	As  noted above, Wolsey~\cite{Wolsey80} observed for the TSP (i.e. for  $s=t$) and a solution $x^*$ to the subtour elimination LP for the TSP that $x^*/2$ is in the $T$-join polyhedron for any set  $ T\subseteq V,\,\hbox{$|T|$  even}$. (It is actually in the form we need it in Cunningham's \cite{Cunningham86}'s independent manuscript.)  If $s\neq t$ this is not true any more for a solution  $x^*$  to the corresponding subtour elimination LP since  for $U\subseteq V$ containing exactly one of $s$ and $t$, $f(U)=1$ in the LP, and hence we are only guaranteed that $x(\delta(U))/2\ge 1/2$. If $U\subset V$ and $|U\cap \{s,t\}|=1$, we call the edge set $\delta(U)$ an {\em $s-t$ cut}. 
	
\begin{figure}
		\begin{center}
			\begin{tikzpicture}[->,shorten >=1pt,auto,node distance=1cm,
			thick,main node/.style={circle,draw,font=\sffamily\bfseries\scriptsize},aux node/.style={}]

			\node[main node] (0) {$s$};

			\foreach \name/\pos in {{1/0}, {2/1}, {4/2}, {5/4}, {6/5}, {t/6}}
			\node[main node] (\name) [right of = \pos] {$\name$};
			
			\foreach \name/\pos in {{Q_1/0},{Q_2/1},{Q_3/2},{Q_4/4},{Q_5/5},{Q_6/6}}
			\node[aux node] (\name) [above right = 1.5cm and 0.1cm of \pos] {$\name$};
			
			\foreach \name/\pos in {{P_1/0},{P_2/1},{P_3/2},{P_4/4},{P_5/5},{P_6/6}}
			\node[aux node] (\name) [below right = 1.5cm and 0.1cm of \pos] {};

			\foreach \name/\pos in {{3/4}}
			\node[main node] (\name) [above of = \pos] {$\name$};
			
			\path[-,every node/.style={font=\sffamily}]
			\foreach \source /\dest in {Q_1/P_1,Q_2/P_2,Q_3/P_3,Q_4/P_4,Q_5/P_5,Q_6/P_6}{
				(\source) edge [very thick,gray!35] node  {} (\dest)
			}
			\foreach \source/\dest in {0/1,4/5,6/t}{
				(\source) edge [very thick, dashed] node {} (\dest)	}
			\foreach \source/\dest in {0/3,1/3,5/2,t/2,3/6}{
				(\source) edge [dotted, bend left] node {} (\dest)	} 			
			(2) edge [dotted] node {} (4)	 			
			\foreach \source/ \dest in {1/2,5/6,3/4} {
				(\source) edge [very thick] node  {} (\dest)
			}
			;
			\end{tikzpicture}
		\end{center}
		\vspace*{-\baselineskip}
		\caption{An example from Gao~\cite{Gao15}: for full edges $x^*(e)=1$, for dashed edges $x^*(e)=2/3$ and for dotted edges  $x^*(e)=1/3$. The narrow cuts are indicated by gray lines and labeled $Q_1,\ldots, Q_6$.} \label{fig:Gao}
	\end{figure}
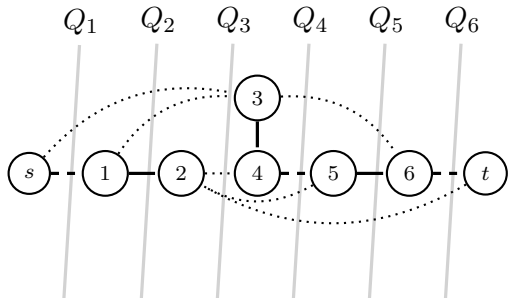

	Following~\cite{AnKS},  we say that a cut $Q$ is {\em narrow} if $x^*(Q) < 2$.  A narrow cut is of course an $s-t$ cut.    We let ${\Qscr}$ be the set of all narrow cuts, that is, ${\Qscr}=\{Q\subset E: \hbox{ $Q$ is an $s-t$ cut, }x^*(Q)<2\}$.  Figure~\ref{fig:Gao} shows an example of an optimal solution $x^*$ to the subtour elimination LP for an $s-t$ path TSP and the narrow cuts $\Qscr$. An, Kleinberg and Shmoys~\cite{AnKS}, observe that  ${\Qscr}$ is defined by a chain of sets of nodes:
	\begin{lemma}[An, Kleinberg, Shmoys~\cite{AnKS}]\label{lem:chain}
		There exist $\{s\}=U_0\subset U_1\subset U_2\subset \ldots \subset U_r=V\setminus\{t\}$ such that \[\Qscr=\{\delta(U_0),\delta(U_1)\ldots, \delta(U_r)\}.\]
	\end{lemma}
	
	For the optimal solution $x^*$ of the subtour elimination LP and the corresponding set of its narrow cuts $\Qscr:=\{\delta(U_0),\delta(U_1),\ldots, \delta(U_r)\}$, Gao's theorem \cite{Gao13} states that there always exists a spanning tree  $S\subseteq \{e:x^*(e)>0\}$ such that $|S\cap Q|=1$ for every narrow cut $Q$. Such a spanning tree will be called a {\em Gao-tree}.  
	
	First note that writing $x^*$ as a convex combination of spanning trees, for all $Q\in\Qscr$ the sum of coefficients of the spanning trees $S$ that have exactly one edge in $Q$ is at least $2-x^*(Q)$ (which was first noted by~\cite{AnKS} and used in all the articles on the subject that followed). 
	Gottschalk and Vygen~\cite{GottschalkV16} generalized Gao's result: a subtour elimination LP solution $x^*$ can be written as a convex combination of ``generalized Gao-trees'', such that  {\em for every $Q\in\Qscr$ simultaneously} the ``first $2-x^*(Q)$ trees" have only a single edge in cut $Q$.
	
	More precisely, let the different cut-sizes of narrow cuts be 
	$2-\lambda_1>2-\lambda_1-\lambda_2>\ldots >2-\lambda_1-\ldots-\lambda_k= 1$, i.e., $\sum_{i=1}^k\lambda_i=1$, and let
	\[\Qscr_i:=\{Q\in\Qscr: x^*(Q)\le 2-\lambda_1-\ldots-\lambda_i\}.\]
	Let $\Bscr_i:=\{ S\subseteq E :\hbox{$S$ is a spanning tree of $G$, $|S\cap Q|=1$ for all }Q\in\Qscr_i\}$, i.e., $\Bscr_1$ is the set of Gao-trees, and $\Bscr_i$ is required to  have a single edge in cuts $Q\in\Qscr_i$ $(i=1,\ldots k)$.  Clearly, if $i<j$, then  $\Bscr_{i}\subseteq \Bscr_j$. Furthermore, define
	\[L_i := \{e\in E: \mbox{there is a unique cut $Q\in\Qscr_i$ such that $e\in Q$}\}.\] Clearly, $\{Q\cap L_i: Q\in\Qscr_i\}$ is a partition of $L_i$, and for each $B\in\Bscr_i$ and $Q\in\Qscr_i$, the unique common edge $e$ of $B$ and $Q$ satisfies $e\in L_i$; then the two components of $B\setminus \{e\}$ are the shores of the cut $Q$. The edges of $L_i$ will be called the {\em lonely edges at layer $i$}. (The lonely edges at layer $1$ are the so called ``Gao-edges''.) We call the cuts in $\Qscr_i$ the {\em lonely cuts at layer $i$}. 
	
	The following simple lemma is fundamental for our algorithm:
	
	\begin{lemma}\label{lem:matroid}
		$\Bscr_i\ne\emptyset$, moreover, $\Bscr_i$ satisfies the basis axioms \cite{Schrijver-book} of a matroid $(i=1,\ldots, k)$.
	\end{lemma} 
	\begin{proof}
		After establishing $\Bscr_i\ne\emptyset$ $(i=1,\ldots k)$, the basis axioms follow from those of spanning trees of a graph and of ``partition matroids'' \cite{Schrijver-book}: for all $B\in\Bscr_i$ the components of $B\setminus L_i$ are spanning trees, so  each $B\setminus L_i$ is a basis in these components, and $B\cap L_i$ consists of exactly one edge in each $Q\in\Qscr_i$. To show that $\Bscr_i\ne\emptyset$ $(i=1,\ldots k)$ we observe that $\Bscr_i \supseteq \Bscr_1$, and it was shown by Gao\cite{Gao13} that $\Bscr_1$ is not empty.	
	\end{proof}

	Denote by $P_i$ the convex hull of $\Bscr_i$  $(i=1,\ldots, k)$. 
	
	\begin{lemma}[\cite{GottschalkV16,SSTvZ}]\label{lem:GV}  
		There exist $x_i\in P_i$ for $i=1,\ldots, k$ such that \[\sum_{i=1}^k\lambda_ix_i=x^*.\]
	\end{lemma}
A statement easily equivalent to this lemma was first proved by Gottschalk and Vygen~\cite{GottschalkV16}; the observation that $P_i$ is the convex hull of bases of a matroid is from \cite{AA}, a connection that led the authors, together with Frans Schalekamp and Vera Traub, to a simple  proof and polynomial algorithm~\cite{SSTvZ}.
	
	\begin{corollary}\label{lemma:treecost} The set of cuts $\Qscr_i$ and minimum-cost spanning tree  $S^*_i$ of $\Bscr_i$ can be computed in strongly polynomial time for $i=1,\ldots, k$, and satisfy
		\[\sum_{i}\lambda_{i}c(S^*_i)\le c(x^*) .\]
	\end{corollary}
	\begin{proof} As noticed by  \cite{Gao13}, given a subtour LP solution $x^*$, the narrow cuts and the different cut- sizes of the narrow cuts can be computed for instance  from  Gomory-Hu trees. 
		This gives the set of cuts $\Qscr_i$ for $i=1,\ldots, k$.
		Finally, the $S^*_i$ $(i=1,\ldots, k)$ can be found using a greedy algorithm in $O(|E|\log|V|)$ time per tree. 
		
		Now use the existence of the convex combination of Lemma~\ref{lem:GV} to conclude:  since $S^*_i$ is a minimum-cost tree in $\Bscr_i$, $c(S^*_i)\le c(x_i)$, and the assertion follows. 
	\end{proof}

	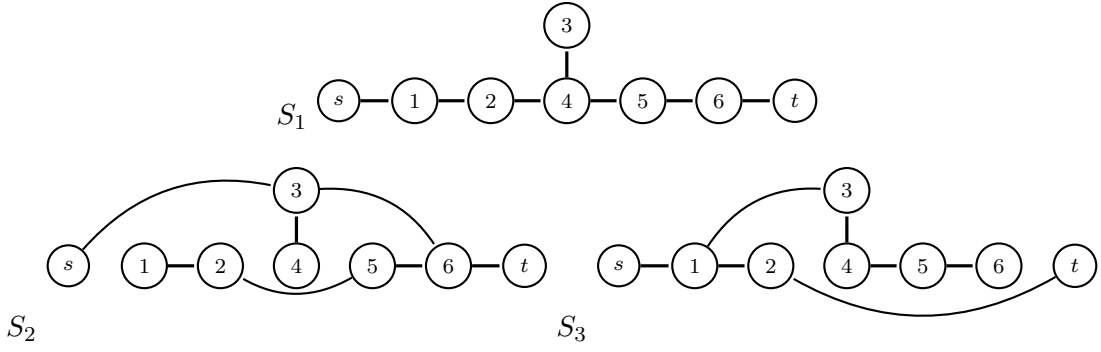
\begin{figure}
		\begin{center}
			$S_1$
			\begin{tikzpicture}[->,shorten >=1pt,auto,node distance=1cm,
			thick,main node/.style={circle,draw,font=\sffamily\bfseries\scriptsize}]

			\node[main node] (0) {$s$};
			
			\foreach \name/\pos in {{1/0}, {2/1}, {4/2}, {5/4}, {6/5}, {t/6}}
			\node[main node] (\name) [right of = \pos] {$\name$};
			
			\foreach \name/\pos in {{3/4}}
			\node[main node] (\name) [above of = \pos] {$\name$};
			
			\path[-,every node/.style={font=\sffamily}]
			\foreach \source/\dest in {0/1,1/2,2/4,4/3,4/5,5/6,6/t}{
				(\source) edge [very thick] node {} (\dest)	}
			;
			\end{tikzpicture}
			\vspace{\baselineskip}
			
			$S_2$
			\begin{tikzpicture}[->,shorten >=1pt,auto,node distance=1cm,
			thick,main node/.style={circle,draw,font=\sffamily\bfseries\scriptsize}]

			\node[main node] (0) {$s$};
			
			\foreach \name/\pos in {{1/0}, {2/1}, {4/2}, {5/4}, {6/5}, {t/6}}
			\node[main node] (\name) [right of = \pos] {$\name$};
			
			\foreach \name/\pos in {{3/4}}
			\node[main node] (\name) [above of = \pos] {$\name$};
			
			\path[-,every node/.style={font=\sffamily}]
			(t) edge [white, bend left] node {} (2)	
			\foreach \source/\dest in {1/2,4/3,5/6,6/t}{
				(\source) edge [very thick] node {} (\dest)	}
			\foreach \source/\dest in {0/3,5/2,3/6}{
				(\source) edge [thick, bend left] node {} (\dest)	} 			
			
			;
			\end{tikzpicture}
			$S_3$
			\begin{tikzpicture}[->,shorten >=1pt,auto,node distance=1cm,
			thick,main node/.style={circle,draw,font=\sffamily\bfseries\scriptsize}]

			\node[main node] (0) {$s$};
			
			\foreach \name/\pos in {{1/0}, {2/1}, {4/2}, {5/4}, {6/5}, {t/6}}
			\node[main node] (\name) [right of = \pos] {$\name$};
			
			\foreach \name/\pos in {{3/4}}
			\node[main node] (\name) [above of = \pos] {$\name$};
			
			\path[-,every node/.style={font=\sffamily}]
			\foreach \source/\dest in {0/1,1/2,4/3,4/5,5/6}{
				(\source) edge [very thick] node {} (\dest)	}
			\foreach \source/\dest in {1/3,t/2}{
				(\source) edge [thick, bend left] node {} (\dest)	} 			
			;
			\end{tikzpicture}
		\end{center}
		
		\caption{Illustration of Lemma~\ref{lem:GV}: $x^*$ from Figure~\ref{fig:Gao} can be expressed as a convex combination of three spanning trees $S_1,S_2,S_3$, each with multiplier $\lambda_{S_i}=1/3$. Note that $S_1$ is a so-called Gao-tree, i.e., $S_1\in \Bscr_1$, and $S_2,S_3\in \Bscr_2$. All edges but $\{3,4\}$ are lonely in $S_1$, and in $S_2$, $S_3$ only the edges incident to $s$ and $t$ are  lonely. }\label{fig:decomp}
	\end{figure}

	\section{The Best-of-Many-with-Deletion (BOMD) Algorithm}\label{sec:alg}
	
	We apply the best-of-many Christofides' algorithm with two modifications. First, rather than expressing $x^*$ as a convex combination of spanning trees and applying Christofides' algorithm to each of the trees of the convex combination, we let $\Sscr^*=\{S^*_1, \ldots, S^*_k\}$ and apply the Christofides' algorithm to the trees in $\Sscr^*$. The second, crucial, difference that leads to an essential improvement is that  we first  {\em delete} the lonely edges, and apply parity correction (and if necessary, reconnection) to the obtained forest.  
	
	More formally, we define
	\[	F^*_i:=S^*_i \setminus L_i.\]
	Note that if we have $|F^*_i\cap Q|=0$ for some narrow cut $Q$, then $Q\in \Qscr_i$.
	
	For $Q\in \Qscr_i$, we denote by $e_i^Q$ the (incidence vector of) the lonely edge of $S^*_i$ in $Q$, i.e., $e_i^Q$ is the unique edge of $S^*_i\cap Q$. 
	By the choice of the greedy algorithm,  $e_i^Q$ is the edge of minimum cost in $Q\cap L_i$.

	Given $F^*_i$, we add a $T_{F^*_i}\tr$-join $J_{F^*_i}$ to $F^*_i$ to obtain a graph in which each node except for $s$ and $t$ has even degree and finally, we add a minimum-cost doubled spanning tree between the components of this graph to reconnect the graph without changing the parity of the degrees. The result  is an $\{s,t\}$-tour.

	If we were to compute a minimum-cost $T_{F^*_i}\tr$-join $J_{F^*_i}$, then $(V, F^*_i+J_{F^*_i})$ may have many components, and the doubled spanning tree between the components could be very expensive. In order to prevent this, we compute a minimum-cost join with respect to a modified cost, which, for every edge $e$, takes the sum of its cost and the anticipation of (or, in fact, an upper bound on) the reconnection cost related to its presence in $J_{F^*_i}$. 
	More precisely, for each edge $e\in E$, let $\Qscr_i(e)$ be the set of cuts in $\Qscr_i$ that contain $e$. Note that $\Qscr_i(e)$ may be empty; in fact, we have $|\Qscr_i(e)|=1$ if $e\in L_i$, $|\Qscr_i(e)|>1$ if $e\in \bigcup_{Q\in \Qscr_i} Q \setminus L_i$ and $|\Qscr_i(e)|=0$ if $e\not \in \bigcup_{Q\in \Qscr_i} Q$.
	
	If, for each $e\in J_{F^*_i}$ we have  $|\Qscr_i(e)|\le 1$, then $(V, F^*_i+J_{F^*_i})$ will be connected, since $J_{F^*_i}$ has at least one edge in every cut $Q\in\Qscr_i$. 
	
	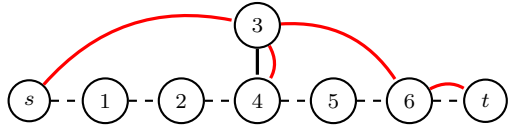
\begin{figure}
		
		\begin{center}
			\begin{tikzpicture}[->,shorten >=1pt,auto,node distance=1cm,
			thick,main node/.style={circle,draw,font=\sffamily\bfseries\scriptsize}]

			\node[main node] (0) {$s$};
			
			\foreach \name/\pos in {{1/0}, {2/1}, {4/2}, {5/4}, {6/5}, {t/6}}
			\node[main node] (\name) [right of = \pos] {$\name$};
			
			\foreach \name/\pos in {{3/4}}
			\node[main node] (\name) [above of = \pos] {$\name$};

			\path[-,every node/.style={font=\sffamily}]
			
			(3) edge [very thick] node {} (4)
			\foreach \source/\dest in {0/3,3/4,3/6,6/t}{
				(\source) edge [red, very thick, bend left] node {} (\dest)	}
			\foreach \source/\dest in {0/1,1/2,2/4,4/5,5/6,6/t}{
				(\source) edge [dashed] node {} (\dest)	}
			
			;
			\end{tikzpicture}
		\end{center}
		
		\caption{The forest $F_1:=S_1\setminus L_1$ obtained by removing the lonely edges from $S_1$ in Figure~\ref{fig:decomp} contains only the edge $\{3,4\}$ indicated in (solid) black, the dashed edges are the lonely edges  of  $S_1$. The (solid) red edges give a possible $T_{F_1}\tr$-join $J_{F_1}$, where edges $\{s,3\}$ and $\{3,6\}$ are bad edges.
			An upper bound on the cost of reconnecting $(V, F_1+J_{F_1})$ can be obtained by adding two copies of two out of the three lonely edges $\{s,1\}, \{1,2\}, \{2,4\}$ and adding two copies of either $\{4,5\}$ or $\{5,6\}$.
		}\label{fig:doubling}
	\end{figure}

	If $|\Qscr_i(e)|\ge 2$, we call $e$ a {\em bad edge} for  layer $i$. See Figure~\ref{fig:doubling} for an illustration. Since lonely edges $e$ for layer $i$ have $|\Qscr_i(e)|=1$, an edge cannot be both bad and lonely. 
	Observe that we can reconnect the forest $(V, F_i^*+ J_{F^*_i})$ by, for each bad edge $b$, adding  $e_i^Q$ for all but one $Q\in\Qscr_i(b)$, and  adding two copies rather than one  to ensure that the degree parities are not changed. The best choice for the unique $Q\in \Qscr_i(b)$,  for which  $e_i^Q$ is omitted instead of being added twice, is of course the one with the largest cost  $c(e_i^Q)$. Therefore, the anticipated reconnection cost of including $e$ in the $T_{F^*_i} \tr$-join will be set to
	\begin{equation}
	r_i(e)= \sum_{Q\in \Qscr_i(e)}2c( e_i^Q) - \max_{Q\in \Qscr_i(e)} 2 c(e_i^Q). \label{eq:reconcost}
	\end{equation}
	Note that $r_i(e)> 0$ only if $|\Qscr_i(e)|\ge 2$, i.e., if $e$ is a bad edge for layer $i$.

	\medskip
	Algorithm~\ref{alg:bomd2} summarizes the description of our algorithm, with a minor simplification:
	
	A {\em spanning forest} of a graph is the union of spanning trees, one spanning tree in  each component. 
	Clearly, taking a minimum cost spanning forest in $(V, E\setminus \bigcup_{Q\in \Qscr_i} Q)$  has exactly the same result as taking a minimum cost spanning tree belonging to $\Bscr_i$, and deleting its lonely edges in $L_i$ $(i=1,\ldots k)$.
	
	\newcommand{\FBT}{\mathrm{Tour}}
	
	\begin{algorithm}[!h]
		\caption{Best-of-Many With Deletion (BOMD) }\label{alg:bomd2}
		{
			Compute an optimal solution $x^*$ to the subtour elimination LP, and the narrow cuts. (\cite{GLS}, \cite{Gao13}).\;
			Let the different cut-sizes of narrow cuts be $2-\lambda_1>2-\lambda_1-\lambda_2>\ldots >2-\lambda_1-\ldots-\lambda_k= 1$.\;
			Let $\Qscr_i$ be the set of narrow cuts $Q$ such that $x(Q)\le 2-\lambda_1-\ldots - \lambda_i$.\;
			\For{$i=1,\ldots, k$}
			{
				Let $F^*_i$ be a minimum-cost spanning forest in $(V, E\setminus \bigcup_{Q\in \Qscr_i} Q)$.\;
				For all $e\in E$, let $r_i(e)$ be defined as in (\ref{eq:reconcost}).\;
				Let $J_{F^*_i}$ be a minimum-cost $T_{F^*_i}\tr$-join with respect to costs $c+r_i$.\;
				Contract the components of $(V, F^*_i+J_{F^*_i})$, and let $2D$ be the edge set of a doubled minimum-cost spanning tree in the contracted graph.\;
				$\FBT_i=F^*_i+J_{F^*_i}+2D$.\;
			}
			Return a minimum-cost $\{s,t\}$-tour from among the set $\bigcup_{i=1,\ldots, k}\{\FBT_i\}$. 
		}
	\end{algorithm}

	We now discuss why the running time of the BOMD algorithm is strongly polynomial. The optimum $x^*$ of the subtour elimination LP can be found in strongly polynomial time \cite[58.5-58.6, pp. 984-986]{Schrijver-book}. The referred results concern the $s=t$ case but the adaptation to arbitrary $s$ and $t$ is obvious, since the minimum size of  a cut separating $s$ and $t$ (which occurs in the modified separation problem) is also strongly polynomial time solvable. 
	As mentioned before, a strongly polynomial time algorithm to list the  narrow cuts $\Qscr$ follows from  \cite{Gao13}, and actually all cuts $Q$ with $x^*(Q)<3$ can be listed \cite{Zenklusen18} in strongly polynomial time, an essential tool for Zenklusen's $3/2$-approximation algorithm.

	Now, from  Lemma~\ref{lemma:treecost} and the fact that a minimum cost $T$-join can be computed in strongly polynomial time \cite{Schrijver-book}, we see that the running time of BOMD is strongly polynomial.  
	
	\section{Analysis}\label{sec:analysis}
	
	In this section, we bound a convex combination of the costs of the constructed $\{s,t\}$-tours, $\sum_{i=1}^k \lambda_i c(\FBT_i)$, by bounding the cost of the forest, parity correction, and reconnection.
	For a quantity $g_i (i=1,\ldots, k)$, we will refer to $\sum_{i=1}^k \lambda_i g_i$ as its {\em average} value. So our goal is to bound the average value of $c(\FBT_i)$.
	
	We begin by observing that, by Lemma~\ref{lemma:treecost}, $\sum_{i=1}^k \lambda_i S_i^*\le c(x^*)$, and that 
	$F^*_i = S_i^* - \sum_{Q\in \Qscr_i}e_i^Q$.

	We define \begin{equation}
	x^{Q} := \sum_{i: Q\in \Qscr_i} \lambda_i e_i^Q.
	\label{xQ}
	\end{equation}
	This is a modification of a vector defined in Seb\H{o}~\cite{Sebo13} for the convex combination $x^*=\sum_{S\in \Sscr: |Q\cap S|=1} \lambda_S (Q \cap S)$; in our definition,  $\Sscr$ is replaced by the greedily found trees $S^*_1,\ldots, S^*_k$, essentially keeping the coefficients, but not $x^*$, and the contribution of  $\lambda_{S^*_i} (Q\cap S_i^*)$ is added to $x^Q$ only if $Q\in \Qscr_i$ (which implies $|Q\cap S^*_i|=1$).

	Using this definition, we have that
	\begin{equation}
	\sum_{i=1}^k \lambda_i F^*_i = \sum_{i=1}^k \lambda_iS^*_i - \sum_{Q\in \Qscr} x^Q.\label{eq:xQtotalsum}
	\end{equation}
	In other words, $\sum_{Q\in \Qscr} x^Q$ gives the average value ``saved'' by deleting edges from the trees $S_1^*, \ldots, S_k^*$.
	
	On the other hand, the $\lambda_i$-value of the trees $S^*_i$ such that $Q$ is lonely is exactly $2-x^*(Q)$, so 
	the average value saved in a specific narrow cut $Q\in \Qscr$ is 
	\begin{equation}
	\sum_{e\in Q}x^Q(e)=\sum_{i:Q\in \Qscr_i} \lambda_i = 2-x^*(Q). \label{eq:xQbound}
	\end{equation}
	
	Intuitively, the idea behind our algorithm and analysis is to use the amount we ``save'' by using $F^*_i$ instead of $S^*_i$ in the layers $i$ such that $Q\in \Qscr_i$, i.e. $x^Q$, to ``pay'' for part of the $T_{F^*_i}\tr$-join $J_{F^*_i}$ and the doubled spanning tree on the components of  $(V, F^*_i+J_{F^*_i})$.

	As explained in Section~\ref{sec:alg}, we can bound the cost of the $T_{F^*_i}\tr$-join $J_{F^*_i}$ and the doubled spanning tree on the components of  $(V, F^*_i+J_{F^*_i})$ by $c(J_{F^*_i})+r_i(J_{F^*_i})$. 
	Since $J_{F^*_i}$ is a minimum-cost $T_{F^*_i}\tr$-join with respect to cost $c+r_i$, we can bound its cost by the cost of a {\em parity correction vector} $y_{i}$ in the $T_{F^*_i}\tr$-join polyhedron. We will refer to $c(y_{i})$ as the cost of parity correction and $r_i(y_i)$ as the cost of reconnection. In the next subsection, we show how to construct these vectors $y_i$ in the $T_{F^*_i}\tr$-join polyhedron for $i=1,\ldots, k$, and in Section~\ref{sec:reconnection}, we analyze the cost of reconnection $r_i(y_i)$ for $i=1,\ldots, k$.

	In Section~\ref{sec:ratio} we summarize the analysis and derive the bound on the approximation ratio (and the integrality gap for the subtour LP) for the BOMD algorithm.
	\subsection{The parity correction vector}\label{sec:parity}

	We construct the parity correction vector in two steps; we first define a {\em basic parity correction at layer $i$}, which will satisfy the constraints of the $T_{F^*_i}\tr$-join polyhedron for cuts that are not narrow. We then add to this vector a {\em parity completion vector} to ensure the vector also satisfies the constraints of the $T_{F^*_i}\tr$-join polyhedron for narrow cuts.

	We use 
	\begin{equation}(1-\gamma)\tfrac12x^*+\gamma S^*_i,\label{eq:BPC}\end{equation}
	as the basic parity correction at layer $i$, where $\gamma\ge 0$ is a parameter whose value will be chosen later. 
	We note that this is essentially the same vector as the one suggested by An, Kleinberg and Shmoys~\cite{AnKS} and later works (even if they use the less intuitive presentation $\beta x^* + \alpha S$, where $S$ is a spanning tree, and  $2\alpha +  \beta\ge 1$).

	Observe that for any cut $Q$ that is not narrow $x^*(Q)\ge 2$ and $S^*(Q) \ge 1$, so $(1-\gamma)\tfrac12x^*(Q) + \gamma S^*(Q) \ge 1$, i.e., the basic parity correction vector is at least 1 across cuts that are not narrow. 
	In order to construct a vector in the $T_{F^*_i}\tr$-join polyhedron, the {\em parity completion vector}, which will add (fractional) edges to the basic parity correction vector for each narrow cut $Q$ if $|Q\cap F^*_i|$ is even, since those are exactly the narrow cuts for which the $T_{F^*_i}\tr$-polyhedron has a constraint.

	For cuts $Q\in \Qscr_i$ we have $|Q\cap F^*_i| = 0$ and $|Q\cap S^*_i|=1$. The basic parity correction vector has a total value of $(1-\gamma)\tfrac12x^*(Q)+\gamma$ on the edges in $Q$, so we need an additional value of $(1-\gamma)\tfrac12(2-x^*(Q))$ as parity completion for such a cut to achieve $y_i(Q)\ge 1$. We will thus add $(1-\gamma)\tfrac12(2-x^*(Q))\, e_i^Q$ to the parity completion vector for the lonely cuts at layer $i$.

	For cuts $Q\in \Qscr\setminus \Qscr_i$ for which $|Q\cap F^*_i|$ is even, we note that $|Q\cap F^*_i|\ge 2$, and thus $|Q\cap S^*_i|\ge 2$. We thus need an additional
	$\left((1-\gamma)\tfrac12(2-x^*(Q)) - \gamma\right)^+$ across every such cut $Q$ to complete the parity correction vector. By (\ref{eq:xQbound}) we have  $x^Q(Q)\ge 2-x^*(Q)$, and thus we can use
	$\left((1-\gamma)\tfrac12- \frac{\gamma}{2-x^*(Q)}\right)^+ x^Q$
	as parity completion for the cuts $Q\in \Qscr\setminus \Qscr^*_i$ for which $|Q\cap F^*_i|$ is even.
	
	Summarizing the above, we have that the parity correction vector at layer $i$ is
	\begin{equation}
	y_i := (1-\gamma)\tfrac12x^*+\gamma\, S^*_i + \sum_{Q\in \Qscr_i} (1-\gamma)\tfrac12(2-x^*(Q))\, e_i^Q 
	+ \sum_{Q\in \Qscr\setminus \Qscr_i}\left((1-\gamma)\tfrac12-\frac{\gamma}{2-x^*(Q)}\right)^+\,x^Q.
	\label{yF}\end{equation}

	Note that we assume in (\ref{yF}) that every cut $Q\in \Qscr$ has an even number of edges in $F^*_i$, and that by (\ref{eq:xQbound}), the vector $y_i$ is in the $T$-join polytope for any $T$ of even size.

	\subsection{Reconnection}\label{sec:reconnection}
	We now analyze $r_i(y_i)$; the cost of the doubled edges that give an upper bound on the cost of the doubled spanning tree for the fractional $T_{F^*_i}\tr$-join $y_i$ defined in (\ref{yF}). 
	Recall that edges $e$ with $r_i(e)>0$ are the {\em bad edges} for $F^*_i$. Equivalently, the bad edges $e$ are those edges that are contained in more than one cut $Q\in \Qscr_i$, i.e., the edges in $\bigcup_{Q\in \Qscr_i} Q \setminus L_i$. 
	Figure~\ref{fig:doubling} shows a forest $F$ and a $T_F\tr$-join $J_F$ with two bad edges.

	\begin{lemma}\label{lemma:reconnection}
		\[r_i(y_i) \le  (1-\gamma)\sum_{Q\in\Qscr_i}(x^*(Q)-1)c(e_i^Q).\]
	\end{lemma}
	\begin{proof}
		In order to show the bound on $r_i(y_i)$, we will need two ingredients. The first ingredient is where we exploit the fact that the set of trees $S_1^*, \ldots, S_k^*$ used by the BOMD algorithm satisfy that $|S_i^*\cap Q|=1$ for all $Q\in \Qscr_i$  ($i=1,\ldots,k$).
		
		\begin{claim}\label{lem:goodpcl} 
			If $e$ is a bad edge for layer $i$, then $y_i(e)=(1-\gamma)\tfrac12x^*(e)$.
		\end{claim}

		\begin{proof_claim}
			Observe that, if $e$ is a bad edge for layer $i$, then $e$ is also a bad edge for layer $h\le i$. Therefore, the trees $S^*_1,\ldots, S^*_i$ do not contain edge $e$; in particular, $e_h^Q\neq e$ for any $h\le i$ and $Q\in \Qscr_h$. To complete the proof, it remains to observe that $y_i-(1-\gamma)\tfrac12x^*$  is non-zero only on edges in $\bigcup_{h\le i}S^*_h$:
			As shown above in (\ref{yF}), $y_i-(1-\gamma)\tfrac12x^*$ consists of $\gamma S^*_i$ plus a parity completion vector that consists of edges $e_i^Q$ for $Q\in \Qscr_i$ and the vectors $x^{Q}$ for $Q\in \Qscr\setminus \Qscr_i$. For $Q\in \Qscr\setminus \Qscr_i$, note that $x^{Q}$ sums up $\lambda_he_h^Q$ for $h$ such that $Q\in \Qscr_h$; since $Q\not\in \Qscr_i$, such $h$ must be strictly smaller than $i$. 
		\end{proof_claim}
		
		The second ingredient is the following claim, which is proved by analyzing the combinatorial structure of the inequalities satisfied by $x$: 
		
		\begin{claim}\label{lem:flow}  
			\[r_i(x^*/2) \le  \sum_{Q_\in\Qscr_i}(x^*(Q)-1)c(e_i^Q).\]
		\end{claim}
		\begin{proof_claim}
			To prove the claim, observe that
			\begin{equation*}
			r_i(e)  = \sum_{Q\in \Qscr_i(e)}2c(e_i^Q) - \max_{Q\in \Qscr_i(e)} 2 c(e_i^Q)\le \sum_{Q\in \Qscr_i(e)}2c(e_i^Q) \left(1-z_{e,Q}\right),\end{equation*}
			for any values $z_{e,Q}\ge 0$ for $e\in E, Q\in \Qscr_i(e)$ such that $\sum_{Q\in \Qscr_i(e)}z_{e,Q}\le 1$.
			Hence, given such values, we can write 
			\begin{eqnarray*}
				r_i(x^*/2)&=& \sum_{e\in E} \frac{x^*(e)}{2}\left(\sum_{Q\in \Qscr_i(e)}2c( e_i^Q) - \max_{Q\in \Qscr_i(e)} 2 c(e_i^Q)\right)\le\sum_{e\in E} \frac{x^*(e)}{2}\left(\sum_{Q\in \Qscr_i(e)}2c( e_i^Q) \left(1-z_{e,Q}\right)\right)\\
				&=&\sum_{Q\in \Qscr_i}c( e_i^Q) \sum_{e\in Q}x^*(e) \left( 1-z_{e,Q}\right)= \sum_{Q\in \Qscr_i} c(e_i^Q) \left(x^*(Q) -\sum_{e\in Q}x^*(e) z_{e,Q}\right).
			\end{eqnarray*}
			
			So in order to prove the claim, it suffices to prove that the following system of inequalities has a solution:
			\begin{eqnarray*}
				\sum_{Q\in \Qscr_i(e)}z_{e,Q}\le 1 && \mbox{ for all }e\in E,\\
				\sum_{e\in Q}x^*(e) z_{e,Q} \ge 1 && \mbox{ for all }Q\in \Qscr_i,\\
				z_{e,Q}\ge 0 && \mbox{ for all }e\in E, Q\in \Qscr_i(e).
			\end{eqnarray*}
			
			By multiplying the first set of inequalities by $x^*(e)$, and letting $f_{e,Q}=x^*(e)z_{e,Q}$, this gives the following problem.
			\begin{eqnarray*}
				\sum_{Q\in \Qscr_i(e)}f_{e,Q}\le x^*(e) && \mbox{ for all }e\in E,\\
				\sum_{e\in Q}f_{e,Q} \ge 1 && \mbox{ for all }Q\in \Qscr_i,\\
				f_{e,Q}\ge 0 && \mbox{ for all }e\in E, Q\in \Qscr_i(e).
			\end{eqnarray*}
			Note that this is in fact a transportation problem with a demand node for each $Q \in\Qscr_i$ with demand 1, 
			a supply node for each $e\in E$ with supply $x^*(e)$, and an arc from $e$ to $Q \in \Qscr_i$ if $e\in Q$, i.e., if $Q\in \Qscr_i(e)$. See Figure~\ref{fig:transportation} for an illustration.
			This transportation problem has a solution $f$ that satisfies all demands if and only for any $\Qscr'\subseteq\Qscr_i$
			\[	x^*(\cup_{Q\in\Qscr'} Q)\ge  |\Qscr'|.
			\]

			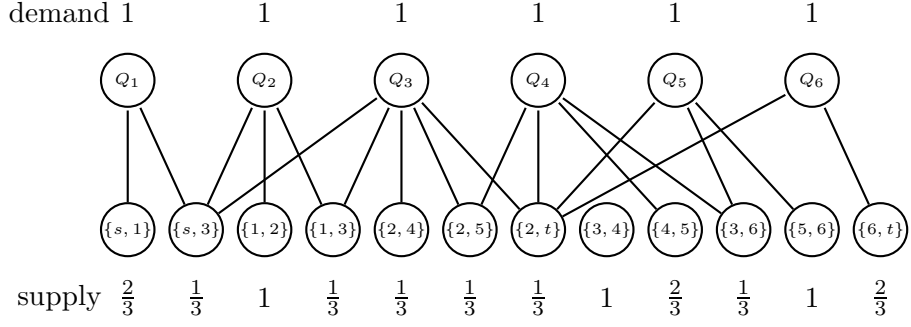
\begin{figure}
				\begin{center}
					\begin{tikzpicture}[->,shorten >=1pt,auto,node distance=0.9cm,
					thick,main node/.style={circle,inner sep = 0pt,minimum size = 20pt,draw,font=\sffamily\bfseries\tiny},aux node/.style={}]

					\node[main node] (s1) {$\{s,1\}$};
					
					\foreach \from/\to/\pos in {{s/3/s1}, {1/2/s3}, {1/3/12}, {2/4/13}, {2/5/24}, {2/t/25}, {3/4/2t},{4/5/34},{3/6/45},{5/6/36},{6/t/56}}
					\node[main node] (\from\to) [right of = \pos] {$\{\from,\to\}$};
					\foreach \pos in {s1,45,6t}
					\node[aux node] (lab\pos) [below of = \pos] {$\frac23$};
					\foreach \pos in {s3,13,24,25,2t,36}
					\node[aux node] (lab\pos) [below of = \pos] {$\frac13$};
					\foreach \pos in {12,56,34}
					\node[aux node] (lab\pos) [below of = \pos] {$1$};	 
					
					\foreach \pos/\c in {s1/1,12/2,24/3,2t/4,45/5,56/6}
					\node[main node] (\c) [above = 1.25cm of \pos] {$Q_\c$};				   		
					\foreach \c in {1,2,3,4,5,6}
					\node[aux node] (lab\c) [above of = \c] {$1$};
					
					\node[aux node] [left of = labs1] {supply};	
					\node[aux node] [left of = lab1] {demand};
					\path[-,every node/.style={font=\sffamily}]	
					foreach \from/\to in {s1/1,s3/1,s3/2,12/2,13/2,s3/3,13/3,24/3,25/3,2t/3,25/4,2t/4,2t/5,2t/6,45/4,36/4,36/5,56/5,6t/6} {
						(\from) edge node  {} (\to)
					};
					\end{tikzpicture}
				\end{center}
				
				\caption{The instance of the transportation problem used to bound the reconnection cost for $S^*_1$ for the example in Figure~\ref{fig:Gao}. There is a node for every edge and every narrow cut $Q\in \Qscr_1=\Qscr$, where each edge-node $e$ has a supply of $x^*(e)$ and every cut-node $Q$ has a demand of 1. } \label{fig:transportation}
			\end{figure}

			Indeed, the necessity and sufficiency of this condition follows from the max flow min cut theorem  (or a variant of the K\H{o}nig-Hall theorem). To prove that the condition is satisfied denote $\ell:=|\Qscr'|$, and let $C_0, \ldots, C_{\ell}$ the node sets of  components of $S\setminus \cup_{Q\in\Qscr'}e_i^Q,$ where $s\in C_0, t\in C_\ell$. Since $\{C_0,\ldots,C_\ell\}$ is a partition of $V$, every edge is counted twice in the sum of their coboundaries:
			\begin{align*}
			x^*(\cup_{Q\in\Qscr'}Q)=\tfrac{1}{2}\sum_{j=0}^\ell x^*(\delta(C_j))&\ge \tfrac 12 \left(x^*(\delta(C_0))+\sum_{j=1}^{\ell-1} x^*(\delta(C_j)) + x^*(\delta(C_\ell))\right)\\
			&\ge \tfrac12 (1+ 2(\ell-1)+1)=\ell.\end{align*}
			exactly as needed. 
		\end{proof_claim}
		
		The lemma now follows directly from Claim~\ref{lem:flow} and the fact that by Claim~\ref{lem:goodpcl}, $r_i(y_i)=r_i(x^*/2)$.
	\end{proof}

	\subsection{Ratio}\label{sec:ratio}
	
	In this section we deduce the ratio that follows from the results of the previous subsections to bound the average of the costs of the constructed $\{s,t\}$-tours, by simply adding up the average cost of the partial results (forest, parity correction, reconnection).

	We show that the parameter $\gamma$ in the basic parity correction vector can be set so that the $x^{Q}$-terms  cancel out.

	\begin{theorem}\label{thm:main}
		The Best-of-Many With Deletion (BOMD) algorithm returns a solution to the $s-t$ path TSP of cost at most $\left(\tfrac32+\frac1{34}\right)\lin$.
	\end{theorem}

	\begin{proof}
		
		We analyze the different parts of the $\{s,t\}$-tours averaged over $S^*_1, S^*_2, \ldots, S^*_k$:

		\noindent {\bf Forest:}
		The average cost of the forests $F^*_1,F^*_2,\ldots, F^*_k$ is 
		\begin{align}
		\sum_{i=1}^k \lambda_i c(F^*_i)&=\sum_{i=1}^k \lambda_i c(S^*_i) - \sum_{Q\in \Qscr} c(x^{Q}) \le c(x^*) - \sum_{Q\in \Qscr} c(x^{Q}), \label{eq:forest}
		\end{align}
		where the equality follows from (\ref{eq:xQtotalsum}) and the inequality follows from Lemma~\ref{lemma:treecost}.

		\noindent {\bf Parity correction:}
		The average basic parity correction vector is 
		\begin{equation}
		\sum_{i=1}^k \lambda_i (1-\gamma)\tfrac12x^* + \sum_{i=1}^k \lambda_k\gamma  S_i^* = 
		(1-\gamma)\tfrac12x^* + \gamma x^* = (1+\gamma) \tfrac12 x^*.\label{eq:basic}\end{equation}
		For the parity completion part of (\ref{yF}), we have that for a given cut $Q\in \Qscr$, the average parity completion is
		\begin{align*}
		&\sum_{i: Q\in \Qscr_i} \lambda_i (1-\gamma)\tfrac12(2-x^*(Q)) \,e_i^Q + \sum_{i: Q\not\in \Qscr_i} \lambda_i \left((1-\gamma)\tfrac12-\frac{\gamma}{2-x^*(Q)} \right)^+ x^Q.
		\end{align*}
		If $(1-\gamma)\tfrac12  -\tfrac{\gamma}{2-x^*(Q)}\ge 0$, we can rewrite this as
		\begin{align}
		&(1-\gamma)\tfrac12(2-x^*(Q)) \sum_{i: Q\in \Qscr_i} \lambda_i e_i^Q + \left((1-\gamma)\tfrac12-\frac{\gamma}{2-x^*(Q)}\right) x^Q \sum_{i: Q\not\in \Qscr_i} \lambda_i \nonumber\\
		&= (1-\gamma)\tfrac12(2-x^*(Q)) x^Q + \left((1-\gamma)\tfrac12-\frac{\gamma}{2-x^*(Q)}\right) x^Q (x^*(Q)-1) \nonumber\\ 
		&= \left((1-\gamma)\tfrac12 - \gamma\frac{x^*(Q)-1}{2-x^*(Q)} \right)x^Q. \label{eq:completion1}
		\end{align}
		where the first equality uses (\ref{xQ}) and (\ref{eq:xQbound}).
		Similarly, if $(1-\gamma)\tfrac12 -\frac{\gamma}{2-x^*(Q)}< 0$, the average parity completion for $Q$ is simply
		\begin{equation}(1-\gamma)\tfrac12(2-x^*(Q)) x^Q.\label{eq:completion2}\end{equation}
		
		\noindent {\bf Reconnection:}
		By Lemma~\ref{lemma:reconnection}, $r_i(y_i) \le \sum_{Q\in \Qscr_i} (x^*(Q)-1) c(e_i^Q)$, and since $\sum_{i: Q\in \Qscr_i}\lambda_i c(e_i^Q) = c(x^{Q})$, we can thus bound the reconnection cost by 
		\begin{equation}\sum_{i=1}^k \lambda_i r_i(y_i)\le\sum_{Q\in \Qscr}(1-\gamma)(x^*(Q) - 1)c(x^{Q}). \label{eq:reconnection}\end{equation}

		\medskip
		\noindent{\bf Total}: 
		If we add up the costs of the bounds on the different parts of the $\{s,t\}$-tours (\ref{eq:forest}), (\ref{eq:basic}), (\ref{eq:completion1},\ref{eq:completion2}), (\ref{eq:reconnection}), we get $(\tfrac32 + \tfrac12 \gamma)c(x^*)$ plus the sum over all $Q\in \Qscr$ of some multiple (that depends on $x^*(Q)$) of $c(x^{Q})$.
		We will show that choosing $\gamma = \frac1{17}$ ensures that the multiplier of $c(x^Q)$ is non-positive for every $Q\in \Qscr$, which implies that the total average cost of the $\{s,t\}$-tours is at most 
		$(\tfrac32 + \tfrac12 \cdot\frac1{17})c(x^*)$.

		If $\tfrac{1}{2}(1-\gamma) - \tfrac{\gamma}{2-x^*(Q)}< 0$, we can use (\ref{eq:completion2}) to bound the parity completion for $Q$, and get a multiplier for $c(x^Q)$ equal to
		\begin{equation*}
		-1 + (1-\gamma)\tfrac12(2-x^*(Q))+ (1-\gamma)(x^*(Q)-1) = (1-\gamma)\tfrac12x^*(Q) -1 < (1-\gamma)-1 <0,
		\end{equation*}
		where the first inequality follows from the fact that $x^*(Q)<2$ if $Q\in \Qscr$.

		If $\tfrac{1}{2}(1-\gamma) - \tfrac{\gamma}{2-x^*(Q)}\ge 0$, , we use (\ref{eq:completion1}) to bound the parity completion for $Q$, and get a multiplier for $c(x^Q)$ equal to
		\begin{equation*}
		-1 + (1-\gamma)\tfrac12 - \gamma\frac{x^*(Q)-1}{2-x^*(Q)} + (1-\gamma)(x^*(Q)-1)= 
		\left(x^*(Q)-\tfrac32\right) - \gamma \left(\left(x^*(Q)-\tfrac32\right) + \frac1{2-x^*(Q)}\right).\end{equation*}
		
		To ensure this multiplier is nonpositive for all $Q\in \Qscr$, we want
		\[\gamma \ge \left(1+\frac{1}{(2-x^*(Q))(x^*(Q)-\tfrac32)}\right)^{-1}. \]
		The minimum value of $\gamma$ for which this is satisfied for all $1<x^*(Q)<2$ is obtained by observing that $(2-x^*(Q))(x^*(Q)-\tfrac32)\le \frac{1}{16}$ (where the maximum is attained for $x^*(Q)=\tfrac74$), which implies that $\gamma= \frac{1}{17}$ indeed ensures that the multiplier of each $x^Q$ is non-positive.		
	\end{proof}

\begin{figure}
		\begin{center}
			\begin{tikzpicture}[->,shorten >=1pt,auto,node distance=1cm,
			thick,main node/.style={circle,draw,font=\sffamily\bfseries\scriptsize},aux node/.style={}]
			
			\node[main node] (0) {$s$};
			\node[main node] (1) [above right = 1.0cm and 1.0cm of 0] {$1$};	
			\node[main node] (2) [below right = 1.0cm and 1.0cm of 0] {$2$};
			\node[main node] (3) [below right = 1.0cm and 1.0cm of 1] {$3$};
			
			\foreach \name/\pos in {{4/3}, {5/4}, {6/5}, {7/6}, {8/7}, {9/8}}
			\node[main node] (\name) [right of = \pos] {$\name$};
			
			\node[main node] (10) [above right = 1.0cm and 1.0cm of 9] {$10$};	
			\node[main node] (11) [below right = 1.0cm and 1.0cm of 9] {$11$};
			\node[main node] (12) [below right = 1.0cm and 1.0cm of 10] {$t$};

			\node[aux node] (Q0) [above right = 2cm and 0.3cm of 0] {};
			\node[aux node] (P0) [below right = 2cm and 0.3cm of 0] {};
			\node[aux node] (Q1) [above right = 2cm and 0.4cm of 0] {};
			\node[aux node] (P1) [below right = 2cm and 1.7cm of 0] {};
			\node[aux node] (Q2) [above right = 2cm and 1.8cm of 0] {};
			\node[aux node] (P2) [below right = 2cm and 1.8cm of 0] {};
			\node[aux node] (Q9) [above right = 2cm and 0.3cm of 9] {};
			\node[aux node] (P9) [below right = 2cm and 0.3cm of 9] {};
			\node[aux node] (Q10) [above right = 2cm and 1.7cm of 9] {};
			\node[aux node] (P10) [below right = 2cm and 0.4cm of 9] {};
			\node[aux node] (Q11) [above right = 2cm and 1.8cm of 9] {};
			\node[aux node] (P11) [below right = 2cm and 1.8cm of 9] {};
			
			\foreach \t in {3,4,5,6,7,8} 
			{
				\node[aux node] (Q\t) [above right = 2cm and 0.2cm of \t] {};
				\node[aux node] (P\t) [below right = 2cm and 0.2cm of \t] {};
			}
			
			\path[-,every node/.style={font=\sffamily}]
			
			\foreach \t in {0,2,3,4,5,6,7,8,9,11}{
				(Q\t) edge [very thick,gray!35] node  {} (P\t)
			}
			
			\foreach \source /\dest in {Q1/P1,Q10/P10}{
				(\source) edge [out= -90, in = 90, very thick,gray!35] node  {} (\dest)
			}	
			
			\foreach \source/\dest in {1/3,9/10}{
				(\source) edge [dotted] node {\tiny{$1/4$}} (\dest)	}
			\foreach \source/\dest in {0/1,10/12}{
				(\source) edge [very thick, dotted] node {\tiny{$3/8$}} (\dest)	}
			\foreach \source/\dest in {2/0,1/2,11/10,12/11}{
				(\source) edge [dashed] node {\tiny{$5/8$}} (\dest)	}
			\foreach \source/\dest in {3/2,11/9,1/10}{
				(\source) edge [very thick, dashed] node  {\tiny{$3/4$}} (\dest)	}
			\foreach \source/\dest in {{4/3}, {5/4}, {6/5}, {7/6}, {8/7}, {9/8}}{
				(\source) edge [very thick] node {} (\dest)	}
			
			;
			\end{tikzpicture}
		\end{center}
		
		\caption{An example on which our arguments are tight. The narrow cuts are indicated by gray lines; each narrow cut except for $(\{s\}, V\setminus\{s\})$ and $(V\setminus\{t\},\{t\})$ have value $7/4$.}\label{fig:badex}
	\end{figure}
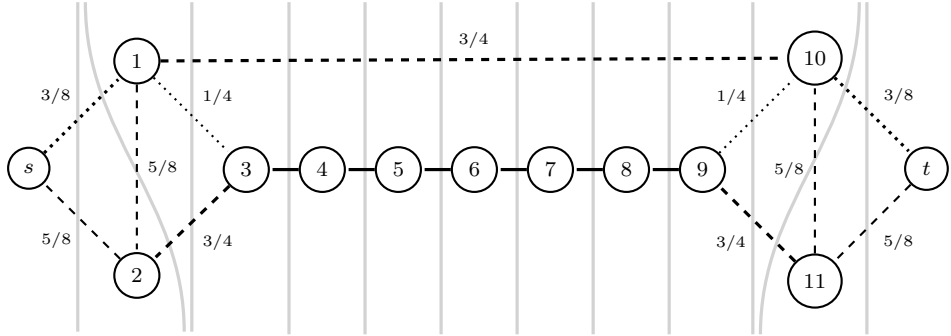

	In Figure~\ref{fig:badex}, we give an example, for which Sylvia Boyd and Frans Schalekamp provided important help,  that shows our arguments are essentially tight. The example consists of a path $P$ of $k=6$ edges, that connect two ``squares''. Every narrow cut (except for the cuts separating only $s$ or only $t$ from the rest of the graph) has value $7/4$, which is the worst case for our analysis. 
	
	In this example, our algorithm would construct two trees $S^*_1,S^*_2$, where $S^*_1$ must be lonely on all narrow cuts (and must thus be an $s-t$ path), and the only requirement on $S^*_2$ is that the degree of $s$ and $t$ is one. Note that all edges of $S^*_1$ thus get deleted to create the $\{s,t\}$-tour based on $S^*_1$, and that edge $\{1,10\}$ is a bad edge for $S^*_1$. So for $S^*_1$, the (bound on the) cost of the resulting $\{s,t\}$-tour is high, since $\{1,10\}$ is used for parity correction with fractional value $\frac12(1-\tfrac{1}{17})\tfrac34$ in our analysis, leading to expensive reconnection in which all but one edge of $P$ is doubled. On the other hand, $S^*_2$ may include both the path $P$ and the edge $\{1,10\}$, which means that parity correction must happen across almost all narrow cuts for $S^*_2$, which is also expensive.
	
	There are a couple of caveats to this bad example, however. First of all, it is not an extreme point of the subtour polytope. Second, in our analysis, we use a fractional solution to bound the cost of parity plus reconnection, which is only tight if the integer solutions that this fractional solution can be decomposed into have the same cost. 
	
	Third, Traub and Vygen~\cite{TraubVpers} assert that analyzing the average cost of the constructed $\{s,t\}$-tours with a modified weighting of the individual tours allows a slightly improved constant of less than $1.5284$. (We note that Traub and Vygen have retracted an initial claim of a bound of less than 1.52.)

	Finally, the example above does not provide a counterexample to the conjecture that $\frac32x^*$ can be expressed as a convex combination of $\{s,t\}$-tours. 

	\bigskip\noindent{\bf Conclusion: } Zenklusen~\cite{Zenklusen18} shows that an $\{s,t\}$-tour of cost at most $1.5$ times the optimum can be found in polynomial time; the conjecture that the integrality gap is $1.5$ is still open.
	There is no counterexample to the conjecture that an $\{s,t\}$-tour of cost at most $1.5$ times the LP bound can be reached with the algorithm we propose (or even the ``vanilla'' best-of-many Christofides' algorithm). The upper bound is indeed,  getting closer and closer to 1.5 for this algorithm. The bottleneck seems the ``reconnection cost'' with a doubled spanning tree for which we found a practical, but not yet ideal treatment.
	
	\bigskip
	\noindent{\bf Acknowledgment} 
	Many thanks are due to  Sylvia Boyd and Frans Schalekamp for helping us to construct the example in Figure~\ref{fig:badex}.

\end{document}